\documentclass[reprint,
superscriptaddress,
amsmath,amssymb,
aps,
prb,
]{revtex4-1}

\usepackage{graphicx}
\usepackage{comment}
\usepackage{dcolumn}
\usepackage{color}
\usepackage{bm}
\usepackage{url}
\usepackage{braket}
\usepackage{amsfonts,amsmath,mathtools}
\usepackage{dsfont}
\usepackage{hyperref}
\usepackage[all]{hypcap}
\setcitestyle{numbers,square}

\newcommand*{\defeq}{\mathrel{\vcenter{\baselineskip0.5ex\lineskiplimit0pt\hbox{\scriptsize.}\hbox{\scriptsize.}}}=}

\begin{document}

\title{Low-frequency behavior of off-diagonal matrix elements in the integrable\\ XXZ chain and in a locally perturbed quantum-chaotic XXZ chain}

\author{Marlon Brenes}
\affiliation{School of Physics, Trinity College Dublin, College Green, Dublin 2, Ireland}
\author{John Goold}
\affiliation{School of Physics, Trinity College Dublin, College Green, Dublin 2, Ireland}
\author{Marcos Rigol}
\affiliation{Department of Physics, The Pennsylvania State University, University Park, PA 16802, USA}

\begin{abstract}
We study the matrix elements of local operators in the eigenstates of the integrable XXZ chain and of the quantum-chaotic model obtained by locally perturbing the XXZ chain with a magnetic impurity. We show that, at frequencies that are polynomially small in the system size, the behavior of the variances of the off-diagonal matrix elements can be starkly different depending on the operator. In the integrable model we find that, as the frequency $\omega\rightarrow0$, the variances are either nonvanishing (generic behavior) or vanishing (for a special class of operators). In the quantum-chaotic model, on the other hand, we find the variances to be nonvanishing as $\omega\rightarrow0$ and to indicate diffusive dynamics. We highlight which properties of the matrix elements of local operators are different between the integrable and quantum-chaotic models independently of the specific operator selected.
\end{abstract}

\maketitle

\section{Introduction}

The eigenstate thermalization hypothesis (ETH)~\cite{Deutsch:1991, Srednicki:1994, Srednicki:1999, Rigol:2008, Alessio:2016} is the paradigm behind our current understanding for why thermalization occurs in generic (quantum-chaotic, nonintegrable) isolated quantum systems and, in particular, in pure states. While pure states remain pure under unitary evolution, i.e., they cannot become any of the mixed states defining traditional ensembles in statistical mechanics (so that thermalization cannot occur at the level of the density matrix of the entire system), observables (few-body operators) can exhibit nontrivial dynamics and equilibration. The matrix elements of observables in the eigenstates of the Hamiltonian, along with the initial state, are the ones that determine the dynamics and expectation values after equilibration. For initial states with subextensive energy fluctuations (which are the ones involved in most experimental situations~\cite{Rigol:2008, Alessio:2016}), it turns out that observables that comply with the ETH are guaranteed to thermalize; that is, after equilibration their expectation values are described by traditional ensembles of statistical mechanics.

Given an observable $\hat O$, the ETH can be written as the following ansatz for the matrix elements $O_{nm}=\langle n|\hat O|m\rangle$ in the energy eigenbasis ($\hat H|m\rangle=E_m|m\rangle$):
\begin{equation}
\label{eq:eth}
O_{n m} = O(\bar{E}) \delta_{n m} + e^{-S(\bar{E}) / 2}f_{O}(\bar{E}, \omega)R_{n m},
\end{equation}
where $\bar{E} \defeq(E_{n} + E_{m}) / 2$ and $\omega \defeq E_{m} - E_{n}$. $S(\bar{E})$ is the thermodynamic entropy at energy $\bar{E}$, $R_{n m}$ is a random variable with zero mean and unit variance, and $O(\bar{E})$ and $f_{O}(\bar{E}, \omega)$ are smooth functions. The first term is the one that ensures that, if the energy fluctuations are subextensive in the initial state, the equilibrated result is described by ensembles of statistical mechanics. The second term ensures that time fluctuations are small at long times so that equilibration occurs (due to the fact that $e^{-S(\bar{E}) / 2}$ is exponentially small in the system size). Many studies have shown that the behavior of the matrix elements of observables in quantum-chaotic systems is described by the ETH ansatz and that, as a result, such systems thermalize under unitary dynamics~\cite{Alessio:2016}.

Integrable systems, on the other hand, are a class known not to thermalize under unitary dynamics for generic (experimentally relevant) initial states~\cite{rigol_dunjko_07, vidmar2016, essler_fagotti_review_16, caux_review_review_16, rigol_16}. Due to the presence of an extensive number of nontrivial conserved quantities, the structure of the matrix elements of observables in integrable systems is different from the one prescribed by the ETH. Two fundamental differences between the behavior of the diagonal matrix elements of observables in integrable and nonintegrable systems at any given energy are that in integrable systems the eigenstate to eigenstate fluctuations do not vanish in the thermodynamic limit~\cite{Rigol:2008, rigol2009breakdown, *rigol_offd_int1, rigol_offd_int2, steinigeweg:2013, beugeling2014finite, vidmar2016, Leblond:2019, Brenes:2020}, while their variance vanishes as a power law in the system size~\cite{biroli2010effect, ikeda2013finite, alba2015, Leblond:2019}, in contrast to the exponential vanishing with system size of the eigenstate to eigenstate fluctuations and their variance in nonintegrable systems~\cite{steinigeweg:2013, Kim_Huse_14, beugeling2014finite, Mondaini:2016, yoshizawa2018numerical, Vidmar_Fabian_19, Leblond:2019}. For the off-diagonal matrix elements of observables, the main difference between interacting integrable systems and quantum-chaotic ones is that in the former the matrix elements are close to lognormally distributed~\cite{Leblond:2019}, while in the latter they are Gaussian distributed~\cite{Moessner:2015, Mondaini:2017, Leblond:2019, Brenes:2020}.

Breaking integrability weakly leads to anomalous unitary dynamics and, specifically, to prethermalization~\cite{moeckel_kehrein_2008, *moeckel_kehrein_2009, eckstein_kollar_09, kollar_wolf_11, tavora_mitra_13, tavora_rosch_14, nessi_iucci_14, essler2014quench, bertini2015prethermalization, *bertini2016prethermalization, fagotti2015universal, lange2018time, reimann_dabelow_19, Mallayya_2019_Prethermalization}, namely, to dynamics that are dictated by the unperturbed integrable Hamiltonian at short times (fast prethermal dynamics), followed by slow thermalizing dynamics dictated by the perturbation~\cite{Mallayya_2019_Prethermalization}. The strength of a global perturbation needed for quantum chaos and thermalization to occur is expected to vanish in the thermodynamic limit~\cite{rabson:2004, santos_rigol_10a, rigol_offd_int2, modak_mukerjee_14a, modak_mukerjee_14b, Pandey:2020}. Remarkably, even a local perturbation such as a magnetic impurity added to an integrable spin chain has been shown to lead to quantum chaos and eigenstate thermalization~\cite{Santos:2004, santos2011domain, torres2014local, Torres_Herrera_2015, XotosIncoherentSIXXZ, Metavitsiadis1, Brenes:2018, Brenes:2020}.

Two recent preprints have explored novel effects of breaking integrability with a local perturbation~\cite{Brenes:2020, Pandey:2020}. In Ref.~\cite{Brenes:2020}, it was shown that the diagonal matrix elements of local operators (with support away from the impurity) satisfy the ETH, with the smooth $O(\bar{E})$ function in Eq.~\eqref{eq:eth} being the microcanonical ensemble predictions for the integrable model, while the off-diagonal matrix elements are Gaussian distributed and comply with the ETH scaling prescribed by Eq.~\eqref{eq:eth}. Remarkably, it was also shown that the variance of the off-diagonal matrix elements of the total spin current operator at low frequencies in the perturbed model exhibits the same ballistic scaling as in the unperturbed integrable model. This is consistent with one of the findings in Ref.~\cite{Brenes:2018}, in which it was shown that transport is ballistic in the quantum-chaotic model. Pandey {\it et al.}~\cite{Pandey:2020}, on the other hand, explored how the norm of the so-called adiabatic gauge potential (the AGP norm) can be used to probe the emergence of quantum chaos. The AGP norm depends on the variance of the off-diagonal matrix elements of the operator used as the perturbation, as well as the energy level spacing, both in the unperturbed Hamiltonian. A remarkable finding in Ref.~\cite{Pandey:2020} is that, depending on the operator chosen to perturb the integrable model, the AGP norm exhibits different scaling with system size. For operators that do not break integrability, the AGP norm scales polynomially, while for operators that break integrability it scales exponentially (as in quantum-chaotic models). Hence, the latter class of operators does not allow one to use the AGP norm to probe the integrability of the unperturbed model.

One of the goals of this work is to explore how the low-frequency behavior of the variance of the off-diagonal matrix elements depends on the operator chosen, at frequencies that are polynomially small in the system size. This allows us to establish a connection between the findings in Refs.~\cite{Brenes:2020} and~\cite{Pandey:2020}. Another goal is to identify which properties of the matrix elements of local operators are different between the integrable and quantum-chaotic model independently of the specific operator selected. Those properties allow one to identify a model as integrable independently of the scaling of the AGP norm.

\section{Hamiltonian and observables}

As in Refs.~\cite{Brenes:2020, Pandey:2020}, our unperturbed integrable model is the XXZ chain with Hamiltonian (we set $\hbar = 1$):
\begin{equation}
\label{eq:h_xxz}
\hat{H}_{\textrm{XXZ}} = \sum_{i=1}^{N-1}\left(\hat{\sigma}^x_{i}\hat{\sigma}^x_{i+1} + \hat{\sigma}^y_{i}\hat{\sigma}^y_{i+1} + \Delta\,\hat{\sigma}^z_{i}\hat{\sigma}^z_{i+1}\right),
\end{equation} 
where $\hat{\sigma}^\nu_{i}$, $\nu = x,y,z$, are the $\nu$-Pauli matrices at site $i$ in a chain with $N$ (even) sites and open boundary conditions. We consider two values of the anisotropy parameter $\Delta$, $\Delta = 0.55$ in the easy-plane regime (as in Ref.~\cite{Brenes:2020}) and $\Delta = 1.1$ in the easy-axis regime (as in Ref.~\cite{Pandey:2020}). We explore the similarities and differences in the behavior of the matrix elements of local operators in those regimes.

We perturb the XXZ chain, placing a magnetic impurity at site $N/2$, which is known to result in a Wigner-Dyson distribution of nearest-neighbor level spacings~\cite{Santos:2004, santos2011domain, torres2014local,  Torres_Herrera_2015, XotosIncoherentSIXXZ, Metavitsiadis1, Brenes:2018}. The single-impurity Hamiltonian reads
\begin{equation}
\label{eq:h_si}
\hat{H}_{\textrm{SI}} = \hat{H}_{\textrm{XXZ}} + h\, \hat{\sigma}^z_{N/2},
\end{equation}
where $h$ (set to $h = 1$) is the impurity field strength. 

The $\hat{H}_{\textrm{XXZ}}$ and $\hat{H}_{\textrm{SI}}$ Hamiltonians commute with the total magnetization $\hat S^z=\sum_i\hat{\sigma}^z_i$, $[\hat{H}_{\textrm{XXZ}}, \hat S^z]=[\hat{H}_{\textrm{SI}}, \hat S^z]=0$. Our calculations are carried out within the zero-magnetization sector, $\braket{\hat{S}^z} = 0$, which is the largest one. In $\hat{H}_{\textrm{SI}}$, reflection symmetry is broken by the impurity. In the calculations involving $\hat{H}_{\textrm{XXZ}}$, we break reflection symmetry by adding a very weak magnetic field at site $i=1$ ($h_1=10^{-1}$). The latter perturbation, like open boundary conditions, does not break integrability~\cite{Santos:2004}. We use full exact diagonalization calculations to compute the matrix elements of observables in the energy eigenbasis. We consider chains with up to $N = 20$ sites, for which the dimension of the Hilbert space of the zero-magnetization sector is $\mathcal{D} = N! / [(N/2)!]^2 = 184\,756$.

We focus on three local operators, the magnetization at site $N/2$, $\hat{\sigma}^z_{N/2}$, which is the operator used to locally perturb the XXZ chain to break its integrability; the next-nearest-neighbor ``kinetic'' energy per site
\begin{equation}
\label{eq:tobsnn}
\hat{T}^\text{NN} \defeq \frac{1}{N} \sum_{i=1}^{N-2} \left( \hat{\sigma}^x_{i}\hat{\sigma}^x_{i+2} + \hat{\sigma}^y_{i}\hat{\sigma}^y_{i+2} \right),
\end{equation}
which, if added to the XXZ Hamiltonian as a perturbation, breaks its integrability (as $\hat{\sigma}^z_{N/2}$ does); and the nearest-neighbor ``kinetic'' energy per site
\begin{equation}
\label{eq:tobs}
\hat{T} \defeq \frac{1}{N} \sum_{i=1}^{N-1} \left( \hat{\sigma}^x_{i}\hat{\sigma}^x_{i+1} + \hat{\sigma}^y_{i}\hat{\sigma}^y_{i+1} \right),
\end{equation}
which, if added to the XXZ Hamiltonian as a perturbation, does not break its integrability (it results in a new XXZ chain). We note that while $\hat{\sigma}^z_{N/2}$ and $\hat{T}^\text{NN}$ are both local intensive operators that, if added as perturbations, break the integrability of $\hat H_\text{XXZ}$, the former has support on a single site (at the center of the chain), while the latter has support on the entire chain, like $\hat T$.

In Ref.~\cite{Pandey:2020}, it was found that the AGP norm of $\hat{\sigma}^z_{N/2}$ at the integrable point scales exponentially with $N$, like the AGP norm of local operators in quantum-chaotic systems. This opens the question of whether $\hat{\sigma}^z_{N/2}$ exhibits any ETH-like behavior at integrability. $\hat{T}$, on the other hand, does not break integrability if added to the XXZ chain, and its AGP norm at the integrable point scales differently from the AGP norm of $\hat{\sigma}^z_{N/2}$~\cite{Pandey:2020}. Here we also study $\hat{T}^\text{NN}$ which, like $\hat{\sigma}^z_{N/2}$, breaks the integrability of $\hat H_\text{XXZ}$ but its support spans over the entire chain.

\section{Diagonal ETH}

\begin{figure}[!t]
\includegraphics[width=\columnwidth]{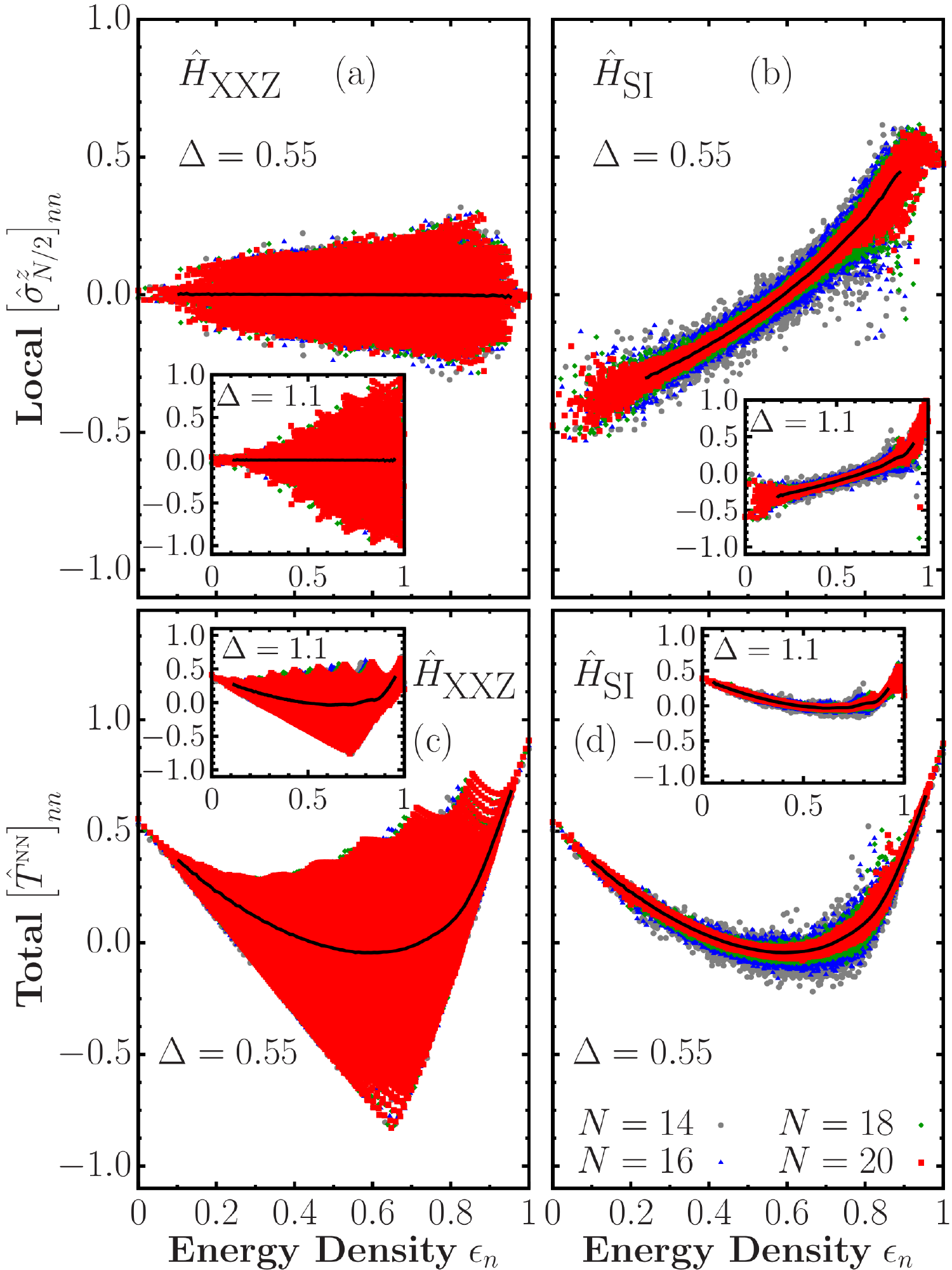}
\vspace{-0.3cm}
 \caption{Diagonal matrix elements of (a) and (b) $\hat{\sigma}^z_{N/2}$ and (c) and (d) $\hat{T}^\text{NN}$ in the eigenstates (a) and (c) of the (integrable) XXZ and (b) and (d) of the (nonintegrable) single-impurity models for $\Delta=0.55$ (main panels) and $\Delta=1.1$ (insets) for different chain sizes $N$. The black lines correspond to the microcanonical averages (within windows with $\delta \epsilon_n = 0.008$) for the largest chain ($N = 20$). We plot the matrix elements vs the energy density $\epsilon_{n}$, defined as $\epsilon_{n} \defeq E_{n} - E_{\textrm{min}} / E_{\textrm{max}} - E_{\textrm{min}}$, where $E_{n}$ is the $n$th energy eigenvalue and $E_{\textrm{min}}$ ($E_{\textrm{max}}$) is the ground-state (highest) energy eigenvalue.}
\label{fig:1}
\end{figure}

In Fig.~\ref{fig:1}, we plot the diagonal matrix elements of $\hat{\sigma}^z_{N/2}$ [Figs.~\ref{fig:1}(a) and~\ref{fig:1}(b)] and $\hat{T}^\text{NN}$ [Figs.~\ref{fig:1}(c) and~\ref{fig:1}(d)] in the eigenstates of $\hat{H}_{\textrm{XXZ}}$ [Figs.~\ref{fig:1}(a) and~\ref{fig:1}(c)] and $\hat{H}_{\textrm{SI}}$ [Figs.~\ref{fig:1}(b) and~\ref{fig:1}(d)] for $\Delta=0.55$ (main panels) and $\Delta = 1.1$ (insets). Figures~\ref{fig:1}(a) and~\ref{fig:1}(c) show that there is no diagonal eigenstate thermalization for these observables in the XXZ chain (the support of the eigenstate to eigenstate fluctuations, at any given energy, does not decrease with increasing system size). This is in contrast to the results for the single-impurity model in which the support of the eigenstate to eigenstate fluctuations of both observables, at any given energy away from the edges of the spectrum, decreases with increasing system size. This suggests that diagonal eigenstate thermalization occurs for $\hat{\sigma}^z_{N/2}$ and $\hat{T}^\text{NN}$ in the single-impurity model.

The results for $[\hat{\sigma}^z_{N/2}]_{nn}$ and $[\hat{T}^\text{NN}]_{nn}$ are in qualitative agreement with the results reported for $\hat T$ in Ref.~\cite{Brenes:2020}, suggesting that diagonal eigenstate thermalization occurs for local operators in the single-impurity model but not in the integrable XXZ chain. A difference to be highlighted between the diagonal ETH for $\hat T$ and $\hat{T}^\text{NN}$ versus $\hat{\sigma}^z_{N/2}$ in the single-impurity model is that for $\hat T$ and $\hat{T}^\text{NN}$ the smooth functions $T(E)$ and $T^\text{NN}(E)$, respectively, are the microcanonical predictions for the integrable model (because $\hat T$ and $\hat{T}^\text{NN}$ are an average over the entire chain and the magnetic impurity is a subextensive perturbation~\cite{Brenes:2020}), while this is clearly not the case for the smooth function $\sigma^z_{N/2}(E)$ of $\hat{\sigma}^z_{N/2}$ in Fig.~\ref{fig:1}(b). The latter is expected since $\hat{\sigma}^z_{N/2}$ is the operator used to perturb the XXZ chain. 

\section{Off-diagonal ETH}

Next, we study the off-diagonal matrix elements in the energy eigenbasis. We explore whether they share properties in the XXZ chain with those of the matrix elements of local operators in quantum-chaotic systems.

\begin{figure}[!t]
\includegraphics[width=\columnwidth]{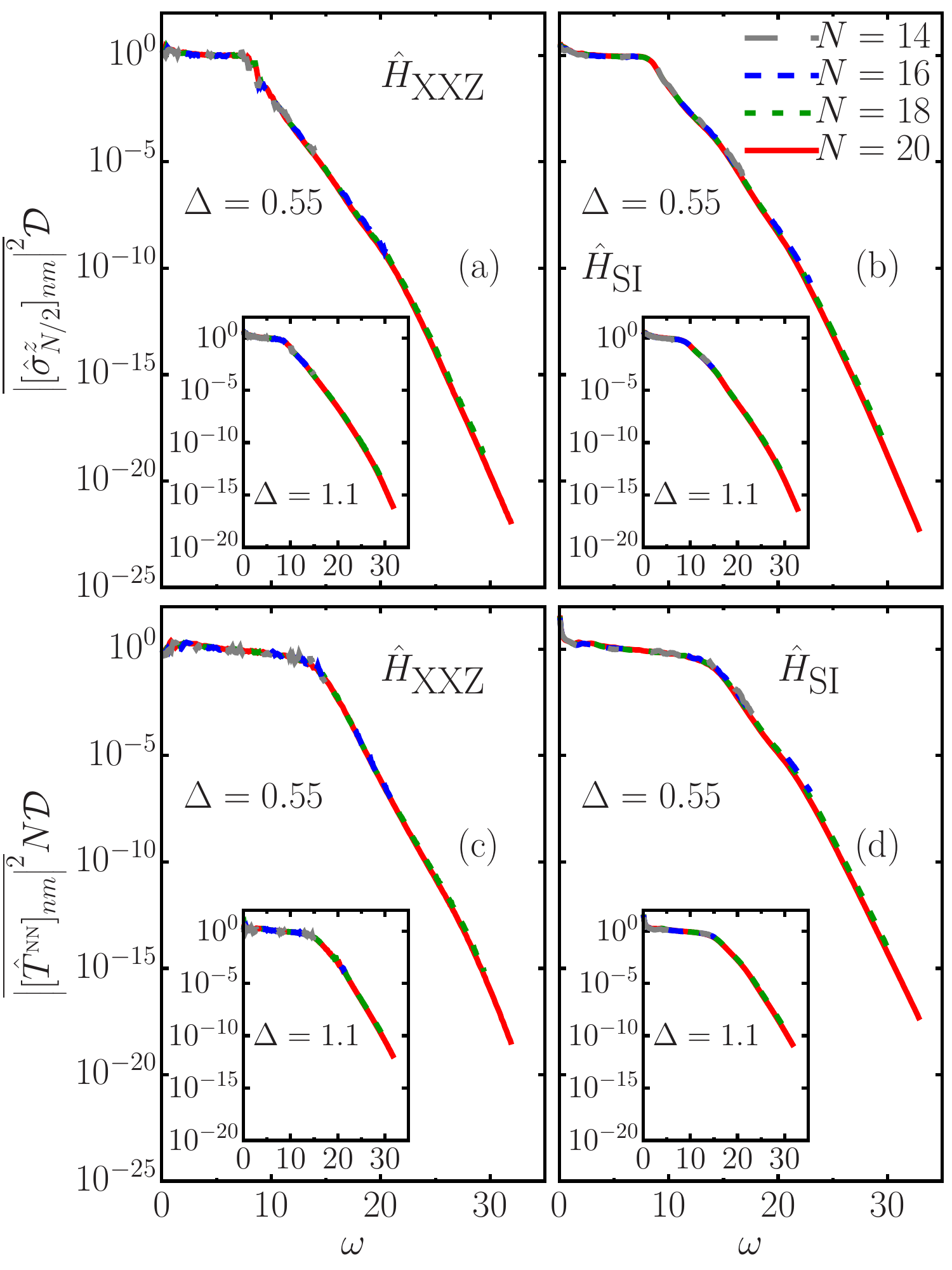}
\vspace{-0.5cm}
 \caption{Average (a) and (b) $\overline{|[\hat{\sigma}^z_{N/2}]_{nm}|^2}$ and (c) and (d) $\overline{|[\hat{T}^\text{NN}]_{nm}|^2}$ vs $\omega$ in the (a) and (c) XXZ and (b) and (d) single-impurity models for $\Delta=0.55$ (main panels) and $\Delta=1.1$ (insets) for different chain sizes $N$. The matrix elements were computed within a small window of energy around $\bar{E} \approx 0$ of width $0.05\epsilon$, where $\epsilon \defeq E_{\textrm{max}} - E_{\textrm{min}}$ denotes the bandwidth. The averages in $\omega$ were calculated in windows with $\delta \omega = 0.1$.}
\label{fig:2}
\end{figure}

In Fig.~\ref{fig:2}, we plot the average $\overline{|[\hat{\sigma}^z_{N/2}]_{nm}|^2}$ [Figs.~\ref{fig:2}(a) and~\ref{fig:2}(b)] and $\overline{|[\hat{T}^\text{NN}]_{nm}|^2}$ [Figs.~\ref{fig:2}(c) and~\ref{fig:2}(d)] vs $\omega$ in the eigenstates of $\hat{H}_{\textrm{XXZ}}$ [Figs.~\ref{fig:2}(a) and~\ref{fig:2}(c)] and $\hat{H}_{\textrm{SI}}$ [Figs.~\ref{fig:2}(b) and~\ref{fig:2}(d)] for $\Delta=0.55$ (main panels) and $\Delta=1.1$ (insets). Since $\overline{[\hat{ \sigma }^z_{ N/2 }]_{nm}}=0$ and $\overline{[\hat{T}^\text{NN}]_{nm}}=0$, the shown averages are the variances of the off-diagonal matrix elements. One can see in Fig.~\ref{fig:2} that the results for $\overline{|[\hat{\sigma}^z_{N/2}]_{nm}|^2}$ and $\overline{|[\hat{T}^\text{NN}]_{nm}|^2}$ are qualitatively (and even quantitatively) similar for the integrable [Figs.~\ref{fig:2}(a) and~\ref{fig:2}(c)] and quantum-chaotic [Figs.~\ref{fig:2}(b) and~\ref{fig:2}(d)] models. For both models, we find the variances to be smooth functions of $\omega$ that decay rapidly at high $\omega$ (the specific scalings with $\omega$, at high $\omega$, are discussed in Ref.~\cite{Leblond:2019}). In Fig.~\ref{fig:2}, we report results for four chain sizes. They exhibit a near-perfect collapse in Figs.~\ref{fig:2}(b) and~\ref{fig:2}(d), showing that in the quantum-chaotic model the variances scale as $1/\mathcal{D}$, as expected from the ETH in the ``infinite-temperature'' regime, namely, when $\bar{E}\approx 0$ and $S(\bar{E}) \simeq \ln{\mathcal{D}}$. The same collapse is seen in Figs.~\ref{fig:2}(a) and~\ref{fig:2}(c). It shows that, as found in Ref.~\cite{Leblond:2019}, the variances exhibit the same scaling in the interacting integrable XXZ model.

The results in Fig.~\ref{fig:2} for the variance of the off-diagonal matrix elements of $\hat{\sigma}^z_{N/2}$ and $\hat{T}^\text{NN}$ in the XXZ and single-impurity models are qualitatively similar between themselves and when compared to the ones reported in Ref.~\cite{Brenes:2020} for $\hat T$. Put together, these results show that within the frequency scales in Fig.~\ref{fig:2}, there are no qualitative differences between the integrable and quantum-chaotic models for the local operators studied. To reveal the existence of qualitative differences between the off-diagonal matrix elements of local operators in integrable and quantum-chaotic systems, at the frequency scales in Fig.~\ref{fig:2}, one needs to study their distributions. The off-diagonal matrix elements of local operators in quantum-chaotic systems are expected to be normally distributed~\cite{Moessner:2015, Mondaini:2017, Leblond:2019, Brenes:2020}, while they were recently found to be close to lognormally distributed in the translationally invariant integrable XXZ chain~\cite{Leblond:2019}.

\begin{figure}[!t]
\includegraphics[width=\columnwidth]{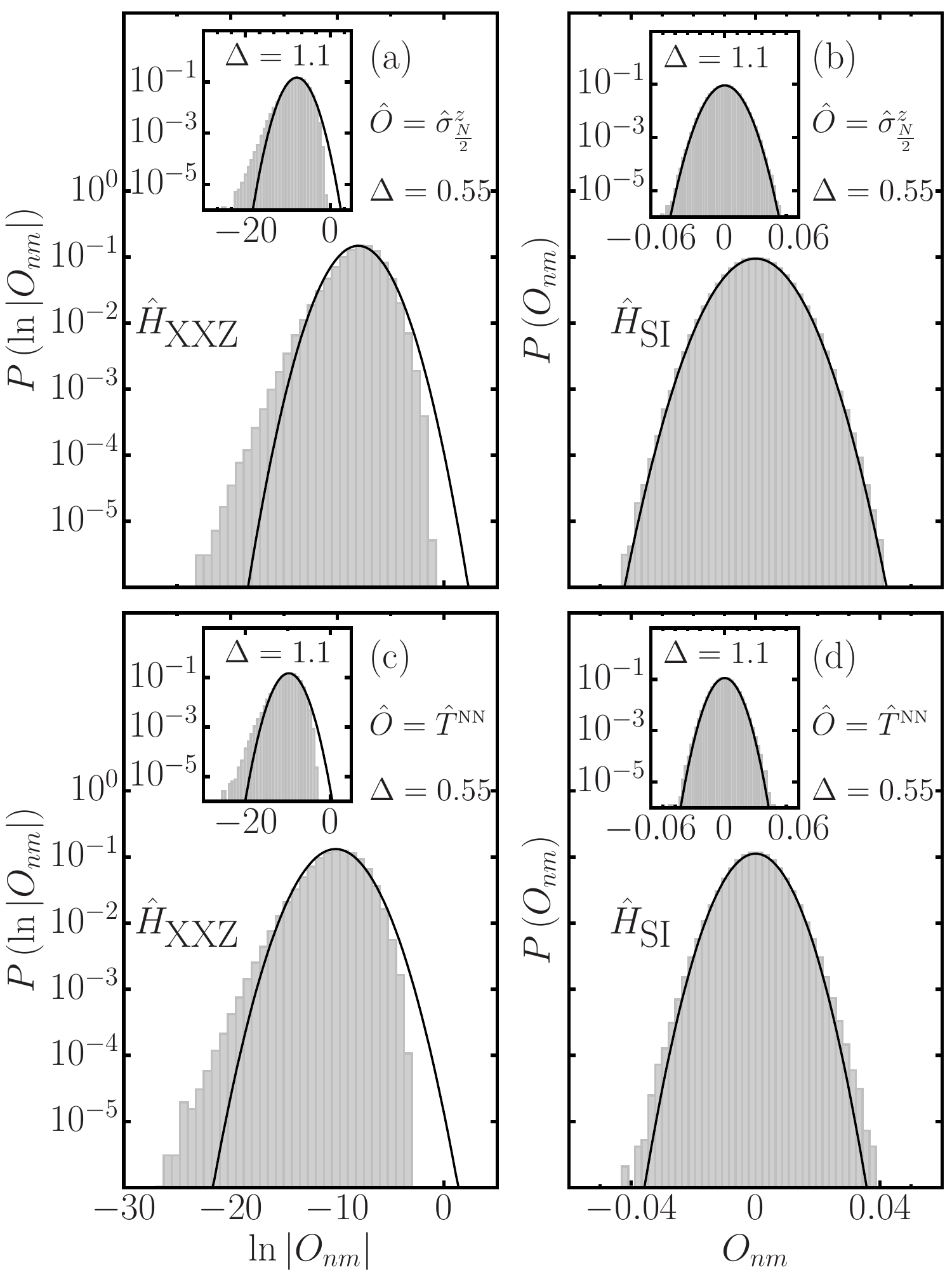}
\vspace{-0.5cm}
\caption{Probability distribution of the off-diagonal matrix elements of (a) and (b) $\hat{\sigma}^z_{N/2}$ and (c) and (d) $\hat{T}^{\textrm{NN}}$ in the eigenstates of (a) and (c) $\hat{H}_{\textrm{XXZ}}$ and (b) and (d) $\hat{H}_{\textrm{SI}}$. The matrix elements used to compute the distributions were selected within a small window of energy around $\bar{E} \approx 0$ of width $0.05\epsilon$, where $\epsilon \defeq E_{\textrm{max}} - E_{\textrm{min}}$ denotes the bandwidth, and $\omega \leq 0.05$.}
\label{fig:3}
\end{figure}

In Fig.~\ref{fig:3}, we show the distributions of off-diagonal matrix elements for $\hat{\sigma}^z_{N/2}$ [Figs.~\ref{fig:3}(a) and~\ref{fig:3}(b)] and $\hat{T}^\text{NN}$ [Figs.~\ref{fig:3}(c) and~\ref{fig:3}(d)] in the eigenstates of $\hat{H}_{\textrm{XXZ}}$ [Figs.~\ref{fig:3}(a) and~\ref{fig:3}(c)] and $\hat{H}_{\textrm{SI}}$ [Figs.~\ref{fig:3}(b) and~\ref{fig:3}(d)] for $\Delta=0.55$ (main panels) and $\Delta=1.1$ (insets). As in Fig.~\ref{fig:2}, we computed these distributions using pairs of eigenstates whose $\bar{E}$ lie in the center of the spectrum, i.e., $\bar{E} \approx 0$, within a small energy window $0.05\epsilon$, where $\epsilon \defeq E_{\textrm{max}} - E_{\textrm{min}}$ denotes the bandwidth. Unlike the results shown in Fig.~\ref{fig:2}, however, we instead focused on the matrix elements with $\omega \leq 0.05$ so that $\omega\approx0$. The results in Figs.~\ref{fig:3}(a) and~\ref{fig:3}(c) are qualitatively similar to the ones reported in Ref.~\cite{Leblond:2019} for translationally invariant observables in the translationally invariant XXZ chain. They show that the off-diagonal matrix elements are nearly lognormally distributed in the eigenstates of $\hat{H}_{\textrm{XXZ}}$ (the skewness of the distributions are similar to the ones in Ref.~\cite{Leblond:2019}). The results in Figs.~\ref{fig:3}(b) and~\ref{fig:3}(d) show that, on the other hand, the off-diagonal matrix elements are normally distributed in the eigenstates of $\hat{H}_{\textrm{SI}}$.

To test whether the off-diagonal matrix elements of $\hat{\sigma}^z_{N/2}$ and $\hat{T}^\text{NN}$ behave for $\omega>0$ as in Fig.~\ref{fig:3} for $\omega\approx0$, we compute~\cite{Leblond:2019}
\begin{equation}
\label{eq:gamma}
\Gamma_{\hat{\sigma}^z_{N/2}}(\omega) \defeq \overline{|[\hat{\sigma}^z_{N/2}]_{nm}|^2} / \overline{|[\hat{\sigma}^z_{N/2}]_{nm}|}^2.
\end{equation}
$\Gamma_{\hat{\sigma}^z_{N/2}} = \pi / 2$ for normally distributed matrix elements.

\begin{figure}[!t]
\includegraphics[width=1.0\columnwidth]{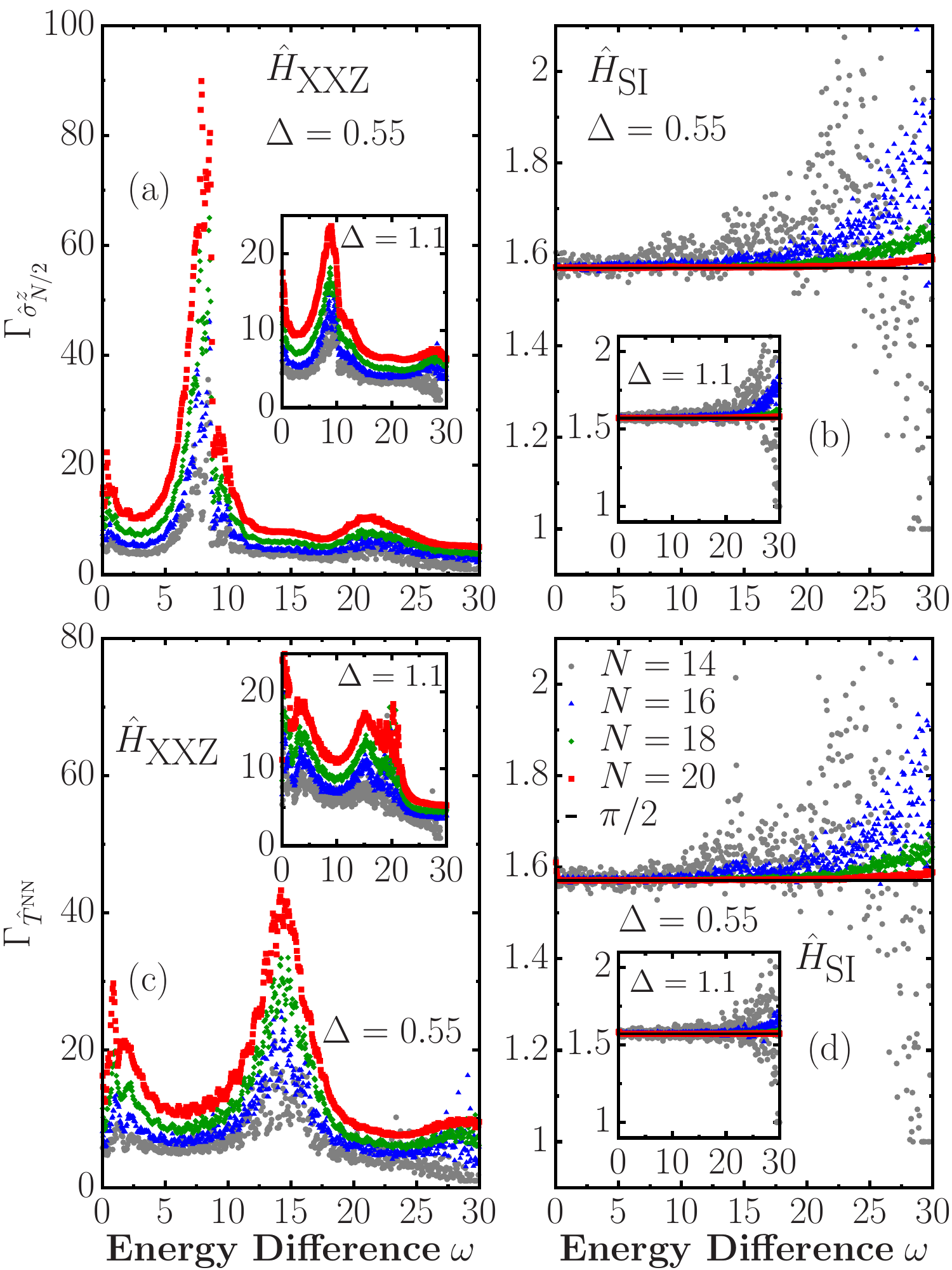}
\caption{(a) and (b) $\Gamma_{\hat{\sigma}^z_{N/2}}$ and (c) and (d) $\Gamma_{\hat{T}^\text{NN}}$ vs $\omega$ [see Eq.~\eqref{eq:gamma}] in the (a) and (c) XXZ and (b) and (d) single-impurity models for $\Delta=0.55$ (main panels) and $\Delta=1.1$ (insets) for different chain sizes $N$. The horizontal line in (b) and (d) marks $\pi/2$. The matrix elements were computed using the same energy window as in Fig.~\ref{fig:2}, while the coarse-graining parameter was chosen to be $\delta \omega = 0.05$.}
\label{fig:4}
\end{figure}

In Fig.~\ref{fig:4}, we plot $\Gamma_{\hat{\sigma}^z_{N/2}}$ [Figs.~\ref{fig:4}(a) and~\ref{fig:4}(b)] and $\Gamma_{\hat{T}^\text{NN}}$ [Figs.~\ref{fig:4}(c) and~\ref{fig:4}(d)] vs $\omega$ in the eigenstates of $\hat{H}_{\textrm{XXZ}}$ [Figs.~\ref{fig:4}(a) and~\ref{fig:4}(c)] and $\hat{H}_{\textrm{SI}}$ [Figs.~\ref{fig:4}(b) and~\ref{fig:4}(d)] for $\Delta=0.55$ (main panels) and $\Delta = 1.1$ (insets). For the quantum-chaotic model [Figs.~\ref{fig:4}(b) and~\ref{fig:4}(d)], we find that the window in $\omega$ within which $\Gamma_{\hat{\sigma}^z_{N/2}}$ and $\Gamma_{\hat{T}^\text{NN}}$, respectively, are essentially $\pi / 2$ increases with increasing system size. We conclude from these results that, for sufficiently large system sizes, the off-diagonal matrix elements of $\hat \sigma^z_{N/2}$ and $\hat{T}^\text{NN}$ in the single-impurity model are normally distributed independently of the value of $\omega$ (as expected for an ETH-satisfying system). On the other hand, for the integrable XXZ chain in Figs.~\ref{fig:4}(a) and~\ref{fig:4}(c), the results for $\Gamma_{\hat{\sigma}^z_{N/2}}$ and $\Gamma_{\hat{T}^\text{NN}}$, respectively, fail to collapse for different systems sizes, showing that the off-diagonal matrix elements of $\hat \sigma^z_{N/2}$ and $\hat{T}^\text{NN}$ are not normally distributed. 

Since the AGP norm for $\hat \sigma^z_{N/2}$ in the XXZ chain scales as in quantum-chaotic models~\cite{Pandey:2020}, the distribution of off-diagonal matrix elements of $\hat \sigma^z_{N/2}$ can be used to identify this model as integrable. We should add that the results in Fig.~\ref{fig:4} for $\Gamma_{\hat{\sigma}^z_{N/2}}$ and $\Gamma_{\hat{T}^\text{NN}}$ are qualitatively similar to the ones reported in Ref.~\cite{Brenes:2020} for $\Gamma_{\hat{T}}$.

\subsection{Variances at low frequency}

\begin{figure}[!t]
\includegraphics[width=\columnwidth]{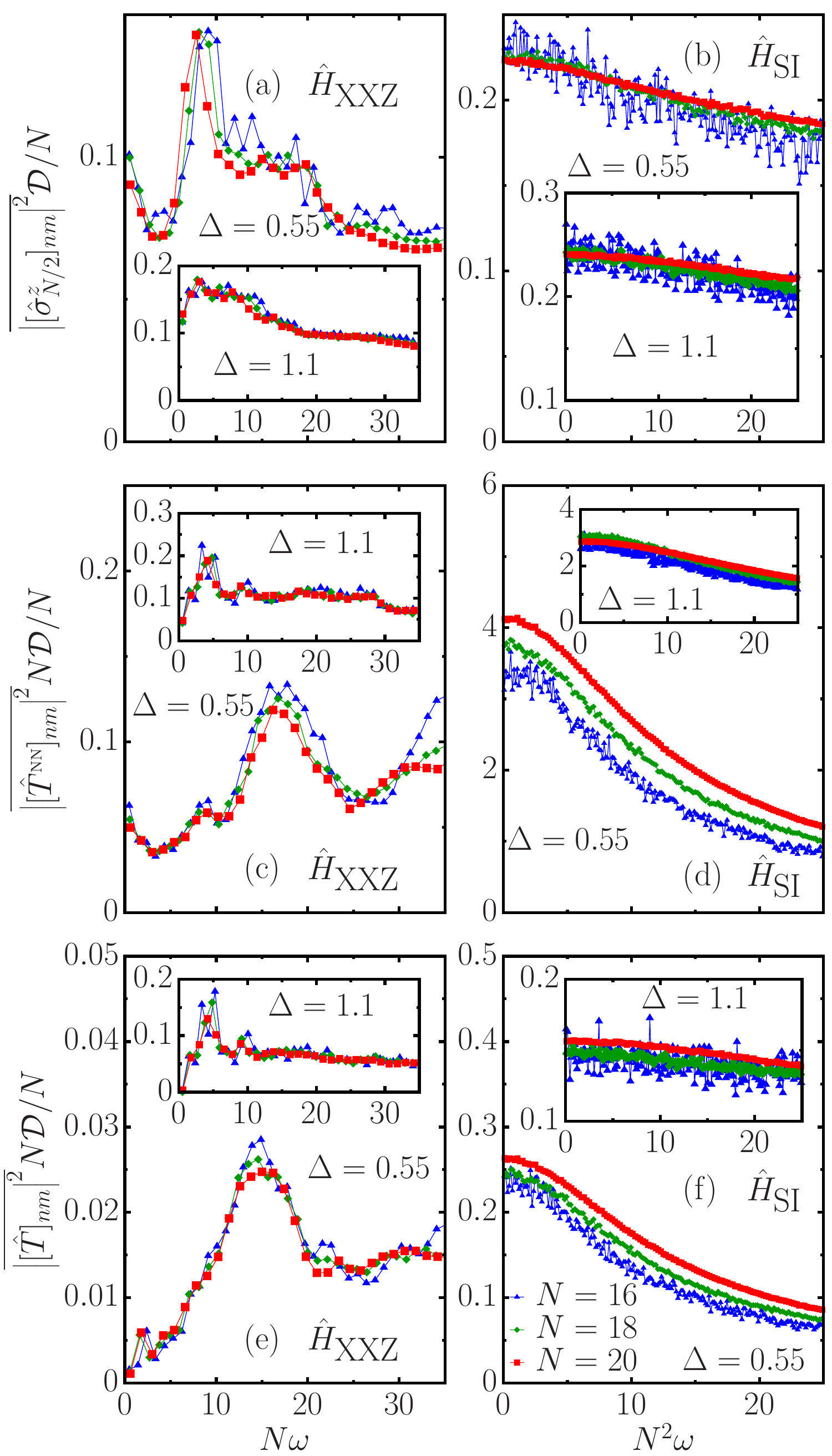}
\vspace{-0.4cm}
\caption{Scaled variances of the off-diagonal matrix elements of (a) and (b) $\hat{\sigma}^z_{N/2}$, (c) and (d) $\hat{T}^\text{NN}$, and (e) and (f) $\hat T$ in the (a), (c), and (e) XXZ and (b), (d), and (f) single-impurity models for different chain sizes $N$. The main panels show results for $\Delta=0.55$, while the insets show results for $\Delta=1.1$. The matrix elements were computed within a small window of energy around $\bar{E} \approx 0$ of width $0.075\varepsilon$. For the binned averages, we used $\delta \omega = 0.06$ for the integrable (left) and $\delta \omega = 5\times10^{-4}$ for the quantum-chaotic (right) models, such that smooth curves are obtained that are robust against changes in $\delta \omega$.}
\label{fig:5}
\end{figure}

Next, we study the low-frequency behavior, at frequencies that are polynomially small frequencies in the system size, of the variances of $\hat \sigma^z_{N/2}$ (which are behind the ``ETH-like'' scaling of the AGP norm in the XXZ chain~\cite{Pandey:2020}), of $\hat T$ (which are behind the ``integrablelike'' scaling of the AGP norm in the XXZ chain~\cite{Pandey:2020}), and of $\hat{T}^\text{NN}$. One of our goals is to identify the distinguishing signatures of integrability and quantum chaos and, within the XXZ chain, of integrability-breaking vs integrability-preserving operators.

In Figs.~\ref{fig:5}(a),~\ref{fig:5}(c), and~\ref{fig:5}(e), we show the scaled variances of the off-diagonal matrix elements of $\hat \sigma^z_{N/2}$, $\hat{T}^\text{NN}$, and $\hat T$, respectively, in the XXZ chain for $\Delta=0.55$ (main panels) and $\Delta=1.1$ (insets). The extra $N$ factor in the $y$ axis in Figs.~\ref{fig:5}(c)--\ref{fig:5}(f) accounts for the Hilbert-Schmidt norm of $\hat T^\text{NN}$ and $\hat T$~\cite{Brenes:2020, Leblond:2019}. For the three observables and both values of $\Delta$, we find a regime in $\omega$ in which the variances indicate ballistic dynamics (they are functions of $N\omega$; see the curves collapse for different system sizes at intermediate values of $N\omega$)~\cite{Alessio:2016}. The collapse degrades as $N\omega$ decreases. This can either be a signature of diffusive dynamics at longer times or just a result of finite-size effects~\footnote{Results for larger chains with periodic boundary conditions suggest diffusive dynamics at longer times~\cite{Leblond:2020}}. Remarkably, a robust feature in Figs.~\ref{fig:5}(a),~\ref{fig:5}(c), and~\ref{fig:5}(e) is that, for both values of $\Delta$ as $\omega\rightarrow0$, the scaled variances of $\hat \sigma^z_{N/2}$ and $\hat{T}^\text{NN}$ do not vanish (they exhibit a peak for $\Delta=0.55$ and a dip for $\Delta=1.1$), while the ones of $\hat T$ vanish. 

Figures~\ref{fig:5}(b),~\ref{fig:5}(d), and~\ref{fig:5}(f), on the other hand, show the striking effect that the single-impurity integrability-breaking perturbation has on the low-frequency behavior of the variances. For the three observables and both values of $\Delta$, the variances become clearly nonvanishing as $\omega\rightarrow0$, and they exhibit a plateau for small values of $N^2 \omega$, indicating diffusive dynamics (finite-size effects appear to affect less the magnitude of the variance of $\hat \sigma^z_{N/2}$ than that of $\hat T^\text{NN}$ and $\hat{T}$), as expected of quantum-chaotic systems~\cite{Alessio:2016}.
 
\section{Summary and discussion}

We showed that the diagonal, and the distribution of the off-diagonal, matrix elements of local operators exhibit distinctive behavior in integrable and quantum-chaotic models independently of the operator chosen. The variances of the off-diagonal matrix elements at intermediate and high frequencies, on the other hand, do not allow one to distinguish integrable from quantum-chaotic models. Instead, one needs to study the low-frequency behavior of the variances to observe differences. For $\omega\rightarrow0$, at frequencies that are polynomially small in the system size, we found the variances to be nonvanishing in the quantum-chaotic model, and in the XXZ chain for two operators that break integrability if added as perturbations to the Hamiltonian ($\hat \sigma^z_{N/2}$ and $\hat{T}^\text{NN}$). On the other hand, in the XXZ chain, we found the variance to vanish for the operator that does not break integrability if added as a perturbation to the Hamiltonian ($\hat T$).

The $\omega\rightarrow0$ behavior observed for $\hat \sigma^z_{N/2}$, $\hat{T}^\text{NN}$, and $\hat T$ in the XXZ chain is consistent with the one inferred from the scaling of the AGP norm in Ref.~\cite{Pandey:2020} at exponentially small frequencies in the system size. Since generic observables are expected to break integrability if added as perturbations to an interacting integrable model, we expect the low-frequency behavior identified here for the variance of the off-diagonal matrix elements of $\hat \sigma^z_{N/2}$ and $\hat{T}^\text{NN}$ to be generic in integrable models (we have already checked that for other observables). This means that using generic observables to compute the AGP norm at integrability does not allow one to identify the model as integrable because the AGP norm would scale as it does for local operators in quantum-chaotic systems. Our results in Fig.~\ref{fig:5} show that, in such situations, studying the distribution of the off-diagonal matrix elements of the operator allows one to distinguish between the two.  

In the quantum-chaotic model generated by the single-impurity perturbation of the XXZ chain, we found that at low frequencies the variances of local operators indicate diffusive dynamics (as expected for generic quantum-chaotic systems~\cite{Alessio:2016}). That said, in Ref.~\cite{Brenes:2020} we showed that the variance of the off-diagonal matrix elements of the total spin current operator for $\Delta=0.55$ in the perturbed model exhibits the same ballistic scaling as in the unperturbed integrable model (consistent with the ballistic nature of spin transport in both models~\cite{Brenes:2018}). All other properties of the matrix elements of the current operator complied with the ETH in the perturbed model but not in the unperturbed integrable one. An interesting open question is whether there are similar exceptions of the lack of low-frequency diffusive scaling in other quantum-chaotic models. 

\acknowledgements
We are grateful to T. LeBlond and L. Vidmar and to the authors of Ref.~\cite{Pandey:2020} for insightful discussions. This work was supported by the European Research Council Starting Grant ODYSSEY Grant No.~758403 (M.B.~and J.G.), the Royal Society (M.B.), a SFI-Royal Society University Research Fellowship (J.G.), and National Science Foundation Grant No.~PHY-1707482 (M.R.). M.B.~and J.G.~acknowledge the DJEI/DES/SFI/HEA Irish Centre for High-End Computing (ICHEC) for the provision of computational facilities and support, project TCPHY118B, and the Trinity Centre for High-Performance Computing. 

\bibliography{bibliography.bib}

\end{document}